\documentclass{article}
\usepackage{hyperref}
\usepackage{amsmath}
\usepackage{amsfonts}
\usepackage{amssymb}
\usepackage{amstext}

\usepackage{graphicx}

\title{Multirate Synchronous Sampling of Sparse Multiband Signals}

\begin{document}

\author{Michael Fleyer, Amir Rosenthal, Alex Linden, and Moshe Horowitz} % <-this % stops a space

\maketitle
\thanks{The authors are with the Technion---Israel Institute of
Technology, Haifa 32000, Israel (e-mail: mikef@tx.technion.ac.il; eeamir@tx.technion.ac.il; alinden@ee.technion.ac.il; horowitz@ee.technion.ac.il).}% <-this % stops a space

\begin{abstract}
Recent advances in optical systems make them ideal for undersampling
multiband signals that have high bandwidths. In this paper we
propose a new scheme for reconstructing multiband sparse signals
using a small number of sampling channels. The scheme, which we call
synchronous multirate sampling (SMRS), entails gathering samples
synchronously at few different rates whose sum is significantly
lower than the Nyquist sampling rate. The signals are reconstructed
by solving a system of linear equations. We have demonstrated an
accurate and robust reconstruction of signals using a small number
of sampling channels that operate at relatively high rates. Sampling
at higher rates increases the signal to noise ratio in samples. The
SMRS scheme enables a significant reduction in the number of
channels required when the sampling rate increases. We have
demonstrated, using only three sampling channels, an accurate
sampling and reconstruction of 4 real signals (8 bands). The
matrices that are used to reconstruct the signals in the SMRS scheme
also have low condition numbers. This indicates that the SMRS scheme
is robust to noise in signals. The success of the SMRS scheme relies
on the assumption that the sampled signals are sparse. As a result
most of the sampled spectrum may be unaliased in at least one of the
sampling channels. This is in contrast to multicoset sampling
schemes in which an alias in one channel is equivalent to an alias
in all channels. We have demonstrated that the SMRS scheme obtains
similar performance using 3 sampling channels and a total sampling
rate 8 times the Landau rate to an implementation of a multicoset
sampling scheme that uses 6 sampling channels with a total sampling
rate of 13 times the Landau rate.
\end{abstract}

\section{Introduction}

In many applications of radars and communications systems it is
desirable to reconstruct a multiband sparse signal from its samples.
When the carrier frequencies of the signal bands are high compared
to the overall signal width, it is not cost effective and often it
is not feasible to sample at the Nyquist rate. It is therefore
desirable to reconstruct the signal from samples taken at rates
lower than the Nyquist rate. There is a vast literature on
reconstructing signals from undersampled data
\cite{Kohlenberg}$-$\cite{Mishali}. Most of methods are based on a
multicoset sampling scheme.

In a multicoset sampling scheme $m$ low-rate cosets are chosen out
of $L$ cosets of samples obtained from time uniformly distributed
samples taken at a rate $F$ where $F$ is greater or equal to the
Nyquist rate $F_{nyq}$ \cite{Venkantaramani}. In each channel the
sampling is offset by a different predetermined integer multiple of
the reciprocal of the rate $F$. The data from the different sampling
channels are then used to reconstruct a signal by solving a system
of linear equations. Under certain conditions on the sampling rate
and number of channels, a proper choice of the time offsets ensures
that the equations have a unique solution in case that the signal
bands are known a priori \cite{Venkantaramani}, or unknown a priori
\cite{Mishali}.

A previous work has demonstrated a different scheme for
reconstructing sparse multiband signals~\cite{Asynchronous}. The
scheme, called multirate sampling (MRS), entails gathering samples
at $P$ different rates. The number $P$ can be small and does not
depend on any characteristics of a signal. The approach is not
intended to obtain the minimum sampling rate. Rather, it is intended
to reconstruct signals accurately with a very high probability at an
overall sampling rate significantly lower than the Nyquist rate
under the constraint of a small number of channels.

The reconstruction method in \cite{Asynchronous} does not require
synchronization of the different sampling channels.  In addition to
reducing the complexity of the sampling hardware, unsynchronized
sampling relaxes the stringent requirement in multicoset sampling
schemes of a very small timing jitter in the sampling time of the
channels. Simulations have indicated that MRS reconstruction is
robust both to different signal types and to relatively high noise.

Accurate signal reconstruction using the unsynchronized MRS scheme
of \cite{Asynchronous} requires that each frequency in the support
of the signal be unaliased in at least one of the sampling channels.
In this paper we describe a new reconstruction algorithm that
overcomes this deficiency by using synchronized sampling channels.
In our synchronized multirate sampling (SMRS) scheme, the aliasing
is resolved, in the same spirit as multicoset sampling schemes, by
solving a system of linear equations. Therefore, the SMRS scheme
enables the reconstruction of signals that cannot be reconstructed
using MRS scheme.

In our SMRS scheme each sampling rate must be an integer multiple of
the same basic frequency. The reconstruction method in our SMRS
scheme requires the same frequency resolution in all sampling
channels despite the fact that the sampling rate is different in
each channel. This requirement is well suited to the implementation
of the scheme using the optical system described in \cite{Avi}. The
sampling is performed in two steps. In the first step the entire
signal spectrum is downconverted into a low frequency region called
baseband by multiplying the signal with a train of short optical
pulses \cite{Avi}. In each of the sampling channels the pulse rate
is different. The frequency downconverted analog signals are then
converted in each sampling channel to digital signals using an A/D
electronic converter that samples at the highest of the channel
rates. The result is that the signal is sampled in all channels with
the same time resolution that is determined by the sampling rate of
the A/D converter. Alternatively, since each sampling rate must be
an integer multiple of the same basic frequency, a common frequency
resolution can be obtained by using the
same time window for all channels. %in all the sampling channels
%while sampling each channel at different sampling rate.
%Alternatively, a common frequency resolution can
%be obtained by using exactly the same time window in all the
%sampling channels. This can be accomplished by choosing all of the
%sampling rates to be an integer multiple of the basic frequency
%increment.

There is an inherent advantage to sampling, in each channel, near
the maximum sampling rate made possible by cost and technology. This
is because sampling at higher rates increases the signal to noise
ratio in sampled signals \cite{Asynchronous}. Our simulation results
indicate that when the sampling rate in each channel increases, our
SMRS scheme requires significantly fewer sampling channels than does
a multicoset sampling scheme of \cite{Mishali} to obtain comparable
reconstruction success. When the sampling rate in each channel
increases, the probability that a sparse signal aliases
simultaneously in all sampling channels becomes very low in an MRS
scheme. It is lower than in a multicoset sampling scheme in which,
because all channels sample at the same frequency, an alias in one
channel is equivalent to an alias in all channels.  Our numerical
simulations indicate that the success rate of our SMRS scheme is
significantly higher than that of a multicoset sampling schemes of
\cite{Mishali} when the number of sampling channels is small (3 in
our simulations) and the sampling rate of of each channel is high.

Exactly the same data obtained in the SMRS scheme can be obtained
using a multicoset sampling scheme. However the multicoset pattern
requires many more channels, each of which samples at a very low
rate. In a numerical
example of section~\ref{Simulations_sect} a 3 channel multirate
sampling pattern is equivalent to a 58 channel multicoset sampling
pattern. With this multicoset pattern the time difference between
two consecutive samples is on the order of 1 psec. Such an accuracy
cannot be practically achieved.

In an SMRS scheme the data is reconstructed differently than in a
multicoset scheme implementation of \cite{Mishali}. In this
multicoset recovery scheme it is assumed that, in addition to being
sparse, the spectrum of the signal consists of bands each of which
is narrower than the coset sampling rate. However, our SMRS scheme
requires no such assumptions on the originating signal. The
reconstruction of sparse signals in our SMRS scheme is robust. This
is because most of the sampled spectrum is unaliased in at least one
of the sampling channels. This is in contrast to the equivalent
multicoset sampling scheme with many low rate sampling channels in
which the alias probability is very high and an alias in one channel
is equivalent to an alias in all channels. In such cases a blind
signal recovery of \cite{Mishali} can be found only by using a
pursuit algorithm whereas, in many cases, the SMRS scheme doesn't
require a pursuit algorithm. The signal
is reconstructed directly by
a single matrix inversion.
By making a very reasonable physical assumption, we are also
able to
simplify the reconstruction by reducing the number of possible signal
locations in a straightforward manner. %The SMRS scheme is also
%expected to be robust even when noise is added during the sampling
%process.

%\textbf{The sampling times can be put on the same time axis. Because
%the sampling rates in our scheme are integer multiples of the
%frequency resolution there is a common time period of the sampling
%times. This is equivalent to the PNS scheme. The sampling times
%within the periodic time window correspond to the different sampling
%channels that sample with the same sampling rate and different time
%offsets. Therefore, our scheme belongs to a class PNS schemes. This
%is similar to a multicoset sampling scheme, except for the fact that
%the sampling times are not a multiple of Nyquist time. An equivalent
%multicoset sampling scheme interpretation can also be obtained by
%finding a common time grid for the SMRS scheme. Then an analysis can
%be performed within a framework of a multicoset sampling scheme of
%\cite{Venkantaramani}. The detailed description of calculating the
%multicoset sampling scheme parameters is provided in the Appendix.
%We show in the numerical example that our scheme can be used to
%efficiently implement multicoset sampling scheme with many number of
%channels. The collected data from low number of channels sampling at
%different rates in our scheme can be used directly for signal
%reconstruction using the algorithm presented below. Alternatively,
%the data can be reordered in time and used with equations from
%\cite{Venkantaramani} or \cite{Mishali} to perform signal
%reconstruction. }

The SMRS scheme reconstructs signals by solving a system of linear
equations. Our simulations indicate that linear equations in an SMRS
scheme are numerically stable in that they have low condition
numbers. This makes them rather insensitive to noise. In the SMRS
scheme, when sampling at total sampling rate that is significantly
higher than theoretically required, the probability that part of the
signal spectrum will not be aliased in at least one of the sampling
channels is very high. The unaliased parts of the signal can be
recovered directly from the sampling channels. Our simulation
results indicate that the sensitivity of the reconstructed signal to
noise added at frequencies of baseband that are not aliased in at
least one of sampling channels is very low.

\section{Synchronized MRS}

In this section we describe the SMRS scheme. Let $F_{\textrm{max}}$
be an assumed maximum carrier frequency and let $X(f)\in
L^2([0,F_{\textrm{max}}])$ be the Fourier transform of a
complex-valued signal $x(t)$ that is to be reconstructed from its
samples. Throughout the analysis we normalize the Fourier transform
by
\[X(f)=\int_{-\infty}^\infty x(t)e^{-j2\pi f t}\;dt.\]
The modifications required to reconstruct real-valued signals are
described in the Appendix. We assume that the signal $x(t)$ to be
sampled, in addition to being bandlimited in $[0,F_{\textrm{max}}]$,
is multiband; i.e., the support of its spectrum is contained within
a finite disjoint union of intervals $(a_n,b_n]$, each of which is
contained in $[0,F_{\textrm{max}}]$. By assumption, $\max{b_n}\le
F_{\textrm{max}}$. In reconstructing a signal we do not assume any a
priori knowledge of the number or location of the intervals
$(a_n,b_n]$. The only requirement is that the signal be sparse;
i.e., that for a signal whose spectral support is contained within
the $N$ intervals $(a_i,b_i]$, $\sum_{k=1}^Nb_k-a_k\ll
F_{\textrm{max}}$.

In the MRS scheme of \cite{Asynchronous} a signal is sampled at $P$
different sampling rates $F^i$ ($i=1\ldots P$). If the delays of the
channels are denoted by $\Delta^i$, the sampled signals $x^i(t)$ are
given by
\begin{equation}\label{sampled spectrum}
x^i(t)=x(t)\sum_{n=-\infty}^\infty\delta\left(t-\frac{n}{F^i}-\Delta^i\right)
\end{equation}
where $\delta(t)$ is a dirac delta ``function''. The spectrum
$X^i(f)$ of the sampled signal in the $i$th channel satisfies
\begin{equation}\label{sampled_spectrum2}
X^i(f)=F^i\sum_{n=-\infty}^\infty
X(f+nF^i)\exp[j2\pi(f+nF^i)\Delta^i].
\end{equation}
Since the channels are synchronized in time, we may set each
$\Delta^i=0$. Equation (\ref{sampled_spectrum2}) becomes
\begin{equation}\label{baseband sum}
X^i(f)=F^i\sum_{n=-\infty}^\infty X(f+nF^i).
\end{equation}

It follows from (\ref{baseband sum}) that all the information about
the $i$th sampled spectrum $X^i(f)$ is contained in the interval
$[0,F^i]$. This interval is called the $i$th baseband. To process
the sampled signals it is necessary to discretize the frequency
axis. We use the same frequency resolution ($\Delta f$) in all of
the sampling channels. The same frequency resolution is directly
obtained if the system is implemented using an optical sampling
system \cite{Avi}. % In such a system the sampling in each sampling
%channel is performed in two steps. In the first step the signal is
%downconverted to baseband by multiplying the signal by a train of
%short optical pulses. In the second step the signal at baseband is
%sampled using a slow electronic analog to digital (A/D) converter.
%In case when the A/D converters in all of the sampling channels
%operate at the same sampling rate (that is greater or equal the
%maximum repetition rate of the the optical pulses used in the
%sampling units) the same frequency resolution is obtained for all of
%the sampling channels.
To use our reconstruction method each sampling rate should be chosen
to be an integer multiple $M^i$ of the basic frequency resolution:
$F^i=M^i\Delta f$.
%The sampling rate of the A/D converter $F_s$ is chosen such that the following
%are satisfied:\begin{enumerate}
%                                                             \item
%                                                             $F_s\geq\max_iF^i$,
%                                                             \item
%                                                             for
%                                                             each
%                                                             $i$,
%                                                             $F_s/F_i$
%                                                             is
%                                                             rational.
%\end{enumerate}
%\noindent
By using a sampling time window with a duration $T=1/\Delta f$, the
same frequency resolution $\Delta f$ is obtained in all of the
sampling channels. Alternatively, the same frequency resolution can
be obtained in an optical system by using A/D converters with the
same rate at the output of all the optical downconvertors
\cite{Avi}.
 %The sampling rate of the $i$th channel is then chosen to be
%$F^i=M^i\Delta f$, where $M^i$ is some integer.
%Alternatively, the same frequency resolution in all of the sampling
%channels $\Delta f$ can be obtained by choosing the sampling rates
%$F^i=M^i \Delta f$ to be an integer multiple $M^i$ of the frequency
%resolution. By using a time window with a duration that is an
%integer multiple of the inverse of each sampling rate it becomes
%possible to work with the same frequency resolution in all of the
%sampling channels.
To reduce computational requirements we ignore
redundant data. Thus, only $M^i$ entries of the Discrete Fourier
transform (DFT) are retained for further processing.
 %We
%use the same spectral resolution $\Delta f$ to represent the
%spectrum of the signal $X(f)$ and the spectra of the $P$ different
%sampled signals $X^i(f)$ ($i=1\ldots P$). Furthermore, the sampling
%rates are chosen to be multiplicities of the spectral resolution,
%i.e. $F^i=M^i \Delta f$, where $M^i$ are positive integers.
%
%The use of a single spectral grid that is consistent with out
%hardware implementation is essential for our reconstruction
%algorithm.
%The reconstruction is subsequently performed on the same grid.

We represent the signals over the common discretized frequency axis
with the following notations:
\begin{eqnarray}\label{discrete}
X^i[k]&=&X^i(k\Delta f),\;\;k=0,  \ldots, M^i-1\nonumber,\\
X[k]&=&X\;(k\Delta f),\;\;k=0,  \ldots, M-1\,\nonumber,
\end{eqnarray}
where $\Delta f$ is the frequency resolution, $M=\lceil F_{\textrm{max}}/\Delta f\rceil$, $X^i[k]$ is the
spectrum of the sampled data from the discretized frequencies in the
baseband $[0,F^i]$ and $X[k]$ is the spectrum of the originating
signal at the discretized frequencies. Equation (\ref{baseband sum})
takes the form
\begin{equation}\label{sum_delt}
    X^i[k]=F^i\sum_{l=0}^{M-1}X[l]\sum^{\infty}_{n=-\infty}\delta[l-(k+nM^i)]
\end{equation}
where $\delta[n]$ is the Kronecker delta function. Equation
(\ref{sum_delt}) can be written in matrix form as follows:
\begin{equation} \label{mat_eq}
\textbf{x}^i=\textbf{A}^i \textbf{x}
\end{equation}
where $\textbf{x}^i$ and $\textbf{x}$ are given by
\begin{align}\label{elements1}
(\textbf{x}^i)_{k+1}&= X^i[k],~~0\leq k\leq M^i-1, \\
\notag (\textbf{x})_{k+1}&=X[k]~~0\leq k\leq M-1,
\end{align}
and $\textbf{A}^i$ is a $M^i\times M$ matrix whose elements are
given by
\begin{equation}\label{basis_func}
\textbf{A}_{k+1,l+1}^i=F^i
\sum^{\infty}_{n=-\infty}\delta[l-(k+nM^i)].
\end{equation}
Each element $\textbf{A}_{k,l}^i$ is equal to either $F^i$ or 0.
This is because there is at most one contribution in the infinite
sum of $\delta$'s which is made when $l\equiv k(\mod M^i)$. In each
row of the matrix $\textbf{A}^i$ there are only $\lfloor
F_{\max}/F^i \rfloor$ non zero elements.

For each value of $i$ ($i=1\ldots P$), (\ref{mat_eq}) defines a set
of linear equations that relate the spectrum of the signal to the
spectrum of its samples. The vector $\textbf{x}$ in (\ref{mat_eq})
is the same for all the $P$ equations because it
doesn't depend on the sampling. %Moreover the number of columns of
%$\underline{\underline{A}}^i$ is the same for all $i$.

The vector $\widehat{\textbf{x}}$ and the matrix
$\widehat{\textbf{A}}$ are obtained by concatenating the vectors
$\textbf{x}^i$ and matrices $\textbf{A}^i$ as follows:
\begin{eqnarray}\nonumber
\widehat{\textbf{x}}= \left(\begin{array}{ccc}
 \textbf{x}^1\\
 \textbf{x}^2\\
\vdots \\
 \textbf{x}^P

\end{array}\right),~
\widehat{\textbf{A}}=
  \left(\begin{array}{ccc}
 \textbf{A}^1\\
\textbf{A}^2\\
\vdots\\
\textbf{A}^P
\end{array}\right).
\end{eqnarray}
These form the system of equations
\begin{equation} \label{mat_eq2}
\widehat{\textbf{x}}=\widehat{\textbf{A}}~ \textbf{x}.
\end{equation}
The matrix $\widehat{\textbf{A}}$ has exactly $P$ non-vanishing
elements in each column that correspond to the locations of the
spectral replica in each channel baseband.

In case that the signal is real-valued its spectrum fulfills
\begin{equation}\label{conjugate}
X(f)=\overline{X}(-f)
\end{equation}
where $\overline{a+bj}=a-bj$ is the complex conjugate. The equations
for reconstructing such a signal are described in the Appendix.

To invert (\ref{mat_eq2}) and calculate the discretized signal
spectrum ($\textbf{x}$) it is necessary that the number of rows
$\sum_{i=1}^P M^i$ in $\textbf{A}$ be equal to or larger than the
number of columns $M$. Defining $F_{\textrm{total}}=\sum_{i=1}^P
F^i$ makes this condition equivalent to the condition
\begin{equation}\label{cond}
F_{\textrm{total}}>F_{\textrm{max}}.
\end{equation}
The condition on the sampling rates given in (\ref{cond}) is
consistent with the requirement that the sampling rate be greater
than the Nyquist rate of a general signal whose spectral support is
$[0,F_{\textrm{max}}]$. However, when sampling sparse signals, an
inversion of the matrix may be possible even if the condition
(\ref{cond}) is not fulfilled. Our objective is to invert
(\ref{mat_eq2}) in the case of sparse signals with sampling rates
$F_{\textrm{total}}<F_{\textrm{max}}$.

\section{Inversion Algorithm}
In this section we describe our inversion algorithm for the SMRS
scheme. The purpose of the algorithm is to invert (\ref{mat_eq2});
i.e., to calculate the vector $\textbf{x}$ from the vector
$\widehat{\textbf{x}}$. To invert the equations with sampling rates
lower than those prescribed by (\ref{cond}) the assumption that the
signal is sparse should be taken into account.

\subsection{Known Band Locations}
In the case in which the signal band locations $(a_n,b_n]$ are known
(\ref{mat_eq2}) can be simplified easily. All the elements of
$\textbf{x}$ that correspond to the frequencies not in the spectral
support $\cup_{n} (a_n,b_n]$ are eliminated from (\ref{mat_eq2}).
All the columns of the matrix which correspond to these elements are
also eliminated. The reduced system of equations that corresponds to
(\ref{mat_eq2}) is given by
\begin{equation} \label{mat_eq4}
\widehat{\textbf{x}}_{\textrm{red}}=\widehat{\textbf{A}}_{\textrm{red}}\textbf{x}_{\textrm{red}}.
\end{equation}

A unique solution exists only if
$\widehat{\textbf{A}}_{\textrm{red}}$ is full column rank. In this
case the inverse can be found using the Moore-Penrose pseudo-inverse
\cite{Penrose}. In a matrix of full column rank the number of rows
must equal or exceed the number of columns.
%This means that the matrix
%$\underline{\underline{\widehat{A}}}_{red}$ must be either
%rectangular or square:
%$\dim\underline{\widehat{X}}_{red}\geq\dim\underline{X}_{red}$.

Although the entire spectrum is downconverted to baseband, we assume
that we are sampling highly sparse signals. Hence the number of
non-zero entries in each $\textbf{x}^i$ is significantly smaller
than $M^i$ in at least one of the sampling channels. The number of
non-zero entries in each $\textbf{x}^i$ might even be smaller
because of aliasing. A necessary condition for a unique inverse or
pseudo-inverse of $\widehat{\textbf{A}}_{\textrm{red}}$ is that this
number still be greater than the number of non-zero entries of
$\textbf{x}$. This is consistent with a Landau theorem \cite{Landau}
that states that one cannot reconstruct a signal if the spectral
density of the samples collected from all sampling channels is less
than the spectral support of the originating signal.

%In order for
%the reduced equations to have a unique solution, the following
%necessary condition must be fulfilled \[\sum_{i=1}^{P} F^i\geq
%\sum_n |b_n-a_n|.\] As a result in Eq.~ref{mat_eq2} the number of
%rows in the matrix is larger than the number of columns. However,
%there may be rows of zeros that correspond to zero amplitude
%frequencies in the baseband, that can be eliminated from the
%equation.

The choice of sampling rates imposes restrictions on the possible
values of $F_{\max}$ for which an inversion of (\ref{mat_eq4}) is
possible. For the matrix $\widehat{\textbf{A}}_{\textrm{red}}$ to
have full column rank, it must not have any identical columns. Since
we do not restrict the possible locations of the known signal bands,
any combination of columns of the matrix $\widehat{\textbf{A}}$ may
appear in the matrix $\widehat{\textbf{A}}_{\textrm{red}}$.
Therefore we require that $\widehat{\textbf{A}}$ not have any
identical columns. The matrix $\widehat{\textbf{A}}$ is composed of
$P$ sub-matrices $\textbf{A}^i$ whose columns are periodic:
\[\textbf{A}_{k,l+M^i}^i=\textbf{A}_{k,l}^i.\]
For the matrix $\widehat{\textbf{A}}$ not to be periodic, it is
required that any common period of the $P$ sub-matrices be larger
than $M$. This condition is met if the least common multiple of the
$\{M^i\}_i$ is larger than $M$. As a result, $F_{\max}$ should
fulfill $F_{\max}<lcm(M^1,\ldots,M^P)\Delta f$, where $lcm$ denotes
least common multiple.

\subsection{Unknown Bands' Location}

In the case that the locations of the bands $(a_n,b_n]$ are not
known a priori some additional assumptions must be made. In the
multicoset recovery scheme of~\cite{Mishali} it was assumed that the
maximum band width is known a priori. In our SMRS scheme we do not
make any assumptions on the intervals $(a_n,b_n]$ but instead we add
assumptions on a signal's spectrum itself.

We assume that, for each discretized frequency $k\Delta f$, any
sampled spectrum $X^i(k\Delta f)=0$ is due only to lack of a signal
in any of its replicas and not due to any aliasing. In other words
if, for any
$n$, $X(f+nF^i)\neq 0$, then $X^i(f)\neq 0$. Another assumption is % Alex thinks it doesn't make sense
that there is a unique solution in the case the signal support is
known, i.e., the matrix $\widehat{\textbf{A}}_{\textrm{red}}$ has a
full column rank for a known support.

Applying the first assumption, one can detect baseband frequencies
in which there is no signal. These frequencies can be eliminated in
the reduction of (\ref{mat_eq2}). We describe this simple procedure
for eliminating frequencies which, according to our assumption,
cannot be part of the spectral support of the originating signal.
The elimination is similar to one presented in asynchronous-MRS
\cite{Asynchronous}. We denote the indicator function $\chi^i[l]$ as
follows:
\begin{equation}\label{indicator}
\chi^i[l]=\left\{
\begin{array}{lll}
1,&&\mathrm{for\;all\;} l\in[0,M-1] \text{\;such that\;}
X^i(l\Delta{f})\neq0\\
0,&&\mathrm{otherwise}.
\end{array}
\right.
\end{equation}
The function $X^i(f)$ is periodic with period $F^i$. Therefore
$\chi^i[l]$ is a periodic extension of an indicator function over
the baseband $f\in[0,F^i)$.

We define the $\chi[l]$ as follows:
\begin{equation}
\chi[l]=\prod_{i=1}^{P}\chi^i[l], ~~~~l\in[0,M-1].
\end{equation}
The function $\chi[l]$ equals 1 over the intersection of all the
upconverted bands of the $P$ sampled signals and it defines the
columns of the matrix $\widehat{\textbf{A}}$ that are retained in
forming the reduced matrix $\widehat{\textbf{A}}_{\textrm{red}}$.
All other columns are eliminated and their corresponding elements in
the vectors $\textbf{x}$ are also eliminated. After the elimination
of the columns from the matrix $\widehat{\textbf{A}}$, the matrix
rows which correspond to zero elements in $\widehat{\textbf{x}}$ and
their corresponding elements in the vectors $\widehat{\textbf{x}}$
are also eliminated. In some cases the function $\chi[l]$ equals 1
only for frequencies within the spectral support of the signal. In
such cases the resulting equations are identical to those found in
the previous subsection (equation (\ref{mat_eq4})). However, in
other cases, $\chi[l]$ may also equal 1 for frequencies outside the
signal's true spectral support. In such cases the reduced matrix
will have more columns than the matrix in the case in which the
spectral support of the signal is known. As a result the inversion
requires finding the values of more variables.

%In the reduction of Eq.~\ref{mat_eq2} to Eq.~\ref{mat_eq4}, for each
%eliminated row there are between $\lfloor F_{max}/\min\{F^i\}\rfloor$
%and $\lceil F_{max}/\max\{F^i\} \rceil$ corresponding columns that
%are eliminated (though some rows may have common corresponding
%columns). Thus, if the number of the zero elements in
%$\widehat{\underline{X}}$ is sufficiently large, the number of rows
%in the matrices $\underline{\widehat{A}}_{red}$ may be larger than
%the number of columns.
Each eliminated zero energy baseband component causes elimination of
respective rows and columns. The elimination of one baseband entry
means that all the frequencies that are downconverted to that
baseband entry (the aliasing frequencies) are also eliminated. This
is because of our first assumption that zero entry in the baseband
corresponds to zero entries in all of the frequency components of
the original signal that are down-converted to frequency of the
baseband entry. Therefore, elimination of one baseband entry results
in elimination of $\lfloor F_{\max}/\min\{F^i\}\rfloor$ to $\lceil
F_{\max}/\max\{F^i\} \rceil$ corresponding columns.  Thus, if the
number of the zero elements in $\widehat{\textbf{x}}$ is
sufficiently large, the number of rows in the matrix
$\widehat{\textbf{A}}_{\textrm{red}}$ may be larger than the number
of columns.

If in addition, matrix $\widehat{\textbf{A}}_{\textrm{red}}$ has a
full column rank, the problem is either consistent or
overdetermined. In such cases there is a unique inversion to
(\ref{mat_eq4}) which can be found using the Moore-Penrose
pseudo-inverse. If the matrix is not full column rank, the problem
is underdetermined and the inversion is not unique. A unique
solution in such cases can be obtained either by increasing the
total sampling rate or by adding additional assumptions on the
signal. \label{unknown_bands_locations}

\subsection{Ill-posed cases}
In many cases the matrix in (\ref{mat_eq4}) for unknown band
locations is not full column rank. In these cases there are subsets
of columns in the matrix $\widehat{\textbf{A}}_{\textrm{red}}$ that
are linearly dependent. Using this linear dependence, a solution to
(\ref{mat_eq4}) can be found. However any solution found can be used
to construct an infinite number of solutions to the equation. Thus,
there is no unique inversse to (\ref{mat_eq4}) and the inversion
problem is ill-posed.

To reconstruct a signal in the case in which the inversion problem
is ill-posed we impose an additional assumption on the signal. We
assume that in the case the signal support is unknown and the
problem is ill-posed, among all possible solutions to
(\ref{mat_eq4}), the originating signal is the one that is composed
of the minimum number of bands. This is the signal we attempt to
construct. We also assumed earlier that the reduced matrix which
corresponds to the case in which the signal support is known is
well-posed. In case that this assumption is not fulfilled the
sampled signal does not contain enough information for solving the
problem.
%The following presents an algorithm that attempts to obtain a solution to
%Eq.~\ref{mat_eq4} that is composed of the minimum number of bands. However, we
%do not provide the conditions under which the correct solution is obtained.

Under the three assumptions stated above (the assumption that leads
to matrix reduction, the existence of the unique solution to
(\ref{mat_eq4}) when the signal bands are known, and band-sparsity)
the inversion problem is reduced to finding the solution of
(\ref{mat_eq4}) that is composed of the minimum number of bands. The
problem is NP-complex since we need to test every possible
combination of bands.

The algorithm described here is of lower complexity and its purpose
is to find a solution of (\ref{mat_eq4}) that is composed of the
minimum number of bands without testing all the combinations. The
resulting algorithm attains a lower success rate but decreases the
run-time significantly as compared to an NP-complex algorithm. We do
not provide the conditions under which the correct solution is
obtained.

Our algorithm is based on the $Orthogonal~Matching~Pursuit~(OMP)$
\cite{Elad}. This algorithm belongs to the category of the "Greedy
Search" algorithms. The original OMP algorithm is used to find the
sparsest solution $\textbf{x}$ of underdetermined equations
$\textbf{A}\textbf{x}=\textbf{b}$ \cite{Elad} where $\textbf{A}$ is
an underdetermined matrix. The sparsest solution is the solution
having the smallest norm $\|\textbf{x}\|_0$ where $\|\textbf{x}\|_0$
is the number of non zero elements in the vector $\textbf{x}$. The
original OMP algorithm collects columns of the matrix $\textbf{A}$
iteratively to construct a reduced matrix $\textbf{A}_{\textrm{r}}$.
At each iteration $n$ the column of $\textbf{A}$ which is added to
$\textbf{A}^{n-1}_{\textrm{r}}$ to produce a matrix
$\textbf{A}^n_{\textrm{r}}$ is the column which results in the
smallest residual error $\min_{\textbf{x}}
\|\textbf{b}-\textbf{A}^n_{\textrm{r}}\textbf{x}\|^2_2$ where for
every vector $\textbf{y}$, $\|\textbf{y}\|^2_2=\sum_i y^2_i$. The
iterations are stopped when some threshold $\epsilon$ is achieved.
Sufficient conditions are given for the algorithm to obtain the
correct solution \cite{Elad}.

We denote $\textbf{A}=\widehat{\textbf{A}}_{\textrm{red}}$,
$\textbf{b}=\widehat{\textbf{x}}_{\textrm{red}}$ and
$\textbf{x}=\textbf{x}_{\textrm{red}}$. Since we are seeking the
solution of $\textbf{A}\textbf{x}=\textbf{b}$ with the smallest
number of bands and not the smallest norm $\|\textbf{x}\|_0$, we
modify the OMP algorithm by instead of choosing a single column as
in \cite{Elad} by selecting iteratively blocks of possible
locations. The columns of the matrix $\textbf{A}$ fall into $J$
blocks. Each block is indexed by $j$ and $B_j$ contains the index of
columns of the $j$th block.

%$\{A_j\}_j$ where each block $A_j$ is a set of columns that corresponds to a %NOTATION????
%where each block indexed by $j$ is denoted by a set of columns'
%indexes $B_j$ that corresponds to a single isolated band in the
%support of $\chi[l]$.
Each index set $B_j$ identifies a possible band of the spectral
support of the reconstructed signal. We start the iteration with the
empty set $S^0=\emptyset$ of column indexes, the empty matrix
$\textbf{A}^0_{\textrm{r}}$, and the set $B^0=\bigcup_{j=1}^J B_j$,
so that at $n$th iteration the following holds: $S^n\bigcup
B^n=B^0$. At the $n$th iteration $(n>1)$ the algorithm must decide
which block to add to $\textbf{A}^{n-1}_{\textrm{r}}$. If the index
set $B_j$ is chosen, then $S^n=S^{n-1}\cup B_j$ and
$B^n=B^{n-1}\setminus B_j$. The matrix $\textbf{A}^n_{\textrm{r}}$
is the matrix whose columns are selected from $\textbf{A}$ according
to the indexing set $S^n$.

%Each block of columns, rather than a single column, is added to
%$S^{n-1}$ and to the matrix
%$\underline{\underline{A}}^{n-1}_{\textrm{r}}$.
The block added is the one that produces the smallest residual error
$\epsilon^n=\min_{B_j\in~B^{n-1}}
\min_{\textbf{x}}\|\textbf{b}-\textbf{A}^n_{\textrm{r,j}}\textbf{x}\|^2_2$
where $\textbf{A}^n_{\textrm{r,j}}$ is the matrix obtained by adding
the block indexed by $B_j$ to $\textbf{A}^{n-1}_{\textrm{r}}$. The
algorithm stops when the threshold $\epsilon$ is reached. The
threshold $\epsilon$ is a very small number and reflects upon the
finite numerical precision of the calculations.

%At the $n^{\mathrm{th}}$ iteration $(n>1)$, a block of columns, rather than a
%single column, is added to $S^{n-1}$ to form the blocks of support
%$S^n$.  A corresponding block is added to the matrix
%$\textbf{A}^{n-1}_{\textrm{r}}$; the resulting matrix being
%$\textbf{A}^n_{r}$.

%The modified $OMP$ algorithm is summarized in the following:  %(??? Not clear enough!)
%
%\begin{itemize}
%  \item Preliminaries: Let $B^0=\bigcup{B_j}$ - union of blocks indexes; Set initial support
%  $S^0=\emptyset$
%  \\n$th$ iteration:
%  \item Sweep: compute errors for each block $B_j\in B^{n-1}$;
%  $\underline{\underline{A}}^n_{\textrm{r,j}}=\underline{\underline{A}}_{(\cdot,S^{n-1}\bigcup{B_j})}$;
%                 $\epsilon^n(j)=\min_{\underline{X}}\|\underline{b}-\underline{\underline{A}}^n_{\textrm{r,j}}\underline{X}\|^2_2$
%  using the least squares solution.
%  \item Update support:
%  $j_n=\text{argmin}_j\epsilon^n(j), S^n=S^{n-1}\bigcup{B_{j_n}}$, remove the used block ($j^n$) $B^n=B^{n-1}\setminus B_{j_n}$
%  \item If $\epsilon^n(j)>\epsilon$ repeat the iteration.
%  \\ Subsequent step:
%  \item Calculate the final result
%  $\underline{X}_{sol}=\text{argmin}_{\underline{X}}\|\underline{b}-\underline{\underline{A}}^n_{\textrm{r}}\underline{X}\|^2_2$
%  using the least squares solution.
%  \item Construct $\underline{X}$ from the corresponding elements in
%  $\underline{X}$.
%\end{itemize}

%Although there are other choices, we chose the modified $OMP$
%algorithm just described that favors block-compact solutions that
%minimize
%$\|\underline{b}-\underline{\underline{A}}\underline{X}\|^2_2$.
The algorithm performed well in our simulations. However, there were
cases in which the support of the reconstructed solution did not
contain all the originating bands and cases in which the
reconstructed signal was incorrect even though all the assumptions
on a signals given in section \ref{unknown_bands_locations} were
fulfilled. Only after performing the last step of the algorithm is
it possible to determine the support of the signal. The algorithm
failed primarily for one of two reasons. One of them was due to the
inclusion of a block that reduced the residual error on one hand but
on the other hand, caused a resulting matrix
$\textbf{A}^n_{\textrm{r}}$ to be not full column rank as
hypothesized in our problem (in section
\ref{unknown_bands_locations}). This can happen, for example, when a
block consists of a correct sub-block and erroneous sub-blocks.
Including any erroneous sub-blocks may result in an ill-posed
problem. Another reason for failure was a large dynamic range of the
signals. When reconstructing such signals, correct bands may be
ignored by the algorithm in cases that the energy within the bands
is significantly lower than the energy in other bands.

Sufficient conditions that assure that the algorithm converges to a unique
solution have not yet been determined.

\section{Simulations results}

The ability of the signal reconstruction algorithm to recover
different types of signals was tested. In one set of simulations the
ability of the algorithm to reconstruct multiband complex and
real-valued signals with different spectral supports, shapes, and
band widths that were not known a priori was tested. Additional
simulations were performed in which real-valued multiband signals
were contaminated by additive white noise. Band carrier frequencies
were chosen from a uniform distribution over the maximum support:
0-20 GHz for complex signal and $-20$ to $20$ GHz for real-valued
signals. For each set of simulations we counted the mean rate of
ill-posed cases in which the modified OMP algorithm was used to
recover the signal. Mean times for accurate signal reconstruction
were also recorded. Failures of the reconstruction were either
because one of the initial assumptions was not fulfilled or because
of the failure of the modified OMP algorithm.

In different simulations the width of each band (hence the total
bandwidth of the signal) was varied. The number of bands was always
set equal to 4 for complex signals and to 8 for real-valued signals
(4 positive bands and 4 negative bands). However, the reconstruction
algorithm was unaware of this number. In all the simulations the
frequency resolution was set to 5 MHz. The simulations were
performed on a 2 GHz Core2Duo CPU with 2 GB RAM storage in the
MATLAB 7.0 environment (no special programming was performed to use
both cores).

%In simulations using
%complex signals, all the bands were chosen to have the same width.  Moreover,
%The carrier frequencies were chosen in such a way to assures there were no
%overlaps between different bands. For a real-valued signals we allowed an
%overlap of a quarter of the band-width between the neighboring bands.
%It
%was also shown possible to detect the true signal support by solving
%Eq.~\ref{mat_eq2} using a 5 MHz resolution and to reconstruct the signal using
%a 1 MHz resolution.

\subsection{Ideal multiband signals}

In the case of ideal multiband signals, because of the absence of
energy outside of strictly defined bands, one can expect a perfect
reconstruction. Accordingly, the algorithm was evaluated by a
perfect reconstruction criterion; i.e., a mean difference between
the true and the reconstructed signal spectrum less than $10^{-10}$.
Whenever this error was attained, the reconstruction was deemed to
have been successful.  Otherwise, it was deemed to have failed. The
threshold for the modified OMP was chosen accordingly;
$\epsilon=10^{-20}$.

Simulations were carried out for complex signals to compare the
results to those published using the multicoset sampling recovery
scheme of \cite{Mishali}. The sampling rates were chosen to be 0.95,
1.0 and 1.05 GHz yielding a total sampling rate $F_{\textrm{total}}=
3.0$ GHz.

\begin{figure}[!t]
  \centering
    \includegraphics[width=3.5in]{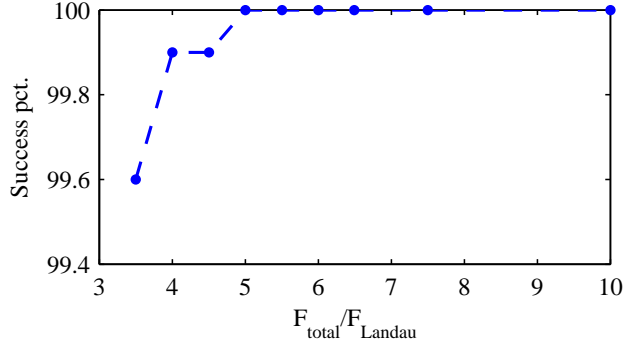}
  \caption{Empirical success percentages for 4-bands complex signals
  for different spectral supports ($F_{\textrm{Landau}}$) with $F_{\textrm{Nyquist}}$=20 GHz,
    total sampling rate $F_{\textrm{total}}=3$ GHz}\label{suc_rate_complex}
\end{figure}

Different signals with 4 bands of equal width were generated. Each
band was chosen to lie within the interval $[0,20]$ GHz. Both the
real and imaginary spectra of the signal within each band were
chosen to be normally distributed. Specifically, for each frequency
$f=k\Delta f$ in a chosen band, the real and imaginary components of
$X(f)$ were chosen randomly and independently from a standard normal
distribution. Each bands' spectra were scaled by a constant $\alpha$
such that each bands' energy was equal to a uniformly generated
value $E$ on the interval $[1,5]$; i.e., for specific band,
\[X(f)=\alpha[X_r(f)+jX_{im}(f)],~\|X(f)\|_2=E.\] These signals were also used to
test the multicoset sampling reconstruction scheme of
\cite{Mishali}. The empirical success rates were obtained from 1000
runs, each with a different total bandwidth ($F_{\textrm{Landau}}$).
The success rate is shown in Fig.~\ref{suc_rate_complex}. We have
validated that the empirical success rate did not significantly
change when the number of simulation runs was increased from 1000 to
5000.

As is evident from Fig.~\ref{suc_rate_complex}, the empirical
success percentage of an ideal reconstruction is high when
$F_{\textrm{total}}/F_{\textrm{Landau}}\ge 5$. In the SBR2 scheme
(downsampling factor $L=199$) in \cite{Mishali} the empirical
success rate shows perfect reconstruction for at least $p=14$
channels. This corresponds to $F_{\textrm{total}}=p/LT=1.4$ GHz and
hence $F_{\textrm{total}}/F_{\textrm{Landau}}$ should be greater
than about 3. Although the total sampling rate in \cite{Mishali} is
lower than in SMRS, the number of channels that are used in that
scheme is significantly higher compared to that used in SMRS where
only 3 channels are used.

Another simulation in~\cite{Mishali} shows that for lower number of
channels with $L=23$ empirical perfect reconstruction is achieved
with at least 6 channels and
$F_{\textrm{total}}/F_{\textrm{Landau}}>13$. In the SMRS scheme
empirical perfect reconstruction was obtained using only three
channels with a total sampling rate
$F_{\textrm{total}}/F_{\textrm{Landau}}\ge 5$. The system parameters
(number of sampling channels, sampling rates, $F_{\max}$) that were
used in our last simulation are the same as those used in our
optical sampling experimental set up. The fact that the simulation
results were obtained in a practical situation demonstrates that our
SMRS scheme can reconstruct sparse signals perfectly using both a
fewer number of sampling channels and with a lower total sampling
rate than are required by multicoset sampling schemes.

We note that the same data that is obtained in a SMRS pattern can
always be obtained by a multicoset sampling pattern since the ratio
between each pair of sampling rates is rational. In our example the
sampling rate of each coset is equal to $1/{LT}=50$ MHz. The number
of multicoset sampling channels ($p$) is 58. The time offset between
the cosets is a multiple of $T=\frac{1}{399~GHz}$. The downsampling
factor $L$ is $399$ GHz$/50$ MHz$=$7980. Note that since $L$ is not
prime, we are not guaranteed to obtain a universal sampling pattern
\cite{Mishali}.

Fig.~\ref{suc_rate_complex} shows that in our scheme 100 percent
empirical success was obtained for
$F_{\textrm{total}}/F_{\textrm{Landau}}>5$. This corresponds to a
maximum overlap in a coset scheme of $K=16$ and hence $p/K=3.6$ for
$p=58$. In comparison, for $p=14$ channels and downsampling factor
of $L=199$ in a multicoset sampling scheme in \cite{Mishali}, the
empirical success rate of 100 percent was obtained for SBR2
algorithm for $p/N=3.5$. This ratio is similar to the equivalent
ratio obtained in our scheme (3.6) although the downsampling factor
in our equivalent multicoset scheme is equal to 7980. For
approximately 95 percent success rate in SBR2 scheme of
\cite{Mishali}, the ratio $p/N$ is 2.75. In our scheme, for the same
empirical success rate of 95 percent,
$F_{\textrm{total}}/F_{\textrm{Landau}}=3.53$ and the equivalent
ratio of the number of channels to maximal number of overlaps is
$p/K=2.4$.

\begin{figure}[!t]
  \centering
    \includegraphics[width=3.5in]{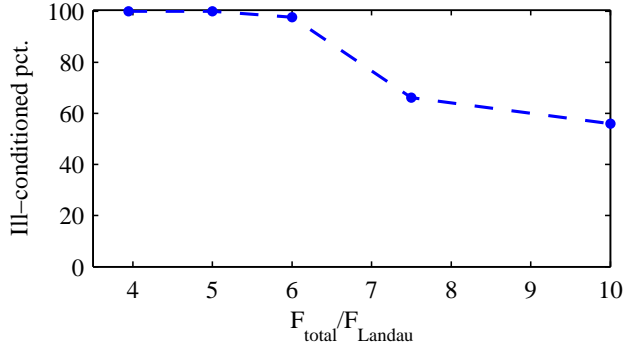}
  \caption{Ill-posed cases mean percentage for 4-bands complex signals for different spectral supports ($F_{\textrm{Landau}}$)
  with $F_{\textrm{Nyquist}}$=20 GHz,
    total sampling rate $F_{\textrm{total}}=3$ GHz}\label{ill_cond}
\end{figure}

The mean percentage of ill-posed cases is shown in
Fig.~\ref{ill_cond}. The figure shows that for $4\leq
F_{\textrm{total}}/F_{\textrm{Landau}} \leq 10$, in most of the
tested cases the inversion was ill posed. Nonetheless, a very high
success percentage was obtained for these values. This indicates
that our modified OMP algorithm was very successful in resolving
these cases.


\begin{figure}[!t]
  \centering
    \includegraphics[width=3.5in]{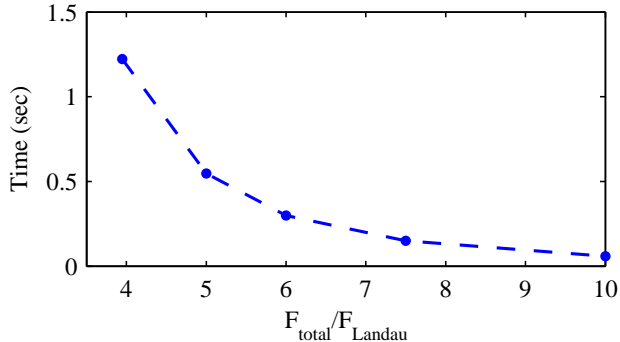}
  \caption{Mean run times for 4-bands complex signals for different
  spectral supports ($F_{\textrm{Landau}}$) with $F_{\textrm{Nyquist}}$=20 GHz,
    total sampling rate $F_{\textrm{total}}=3$ GHz}\label{mean_times_complex}
\end{figure}

Fig.~\ref{mean_times_complex} shows the mean run time as a function
of $F_{\textrm{total}}/F_{\textrm{Landau}}$ (constant total sampling
rate and varying signal support). Because matrix inversion is the
most computationally intensive operation in the algorithm, mean run
time decreases with a reduction in signal bandwidth. This is
because, with a fixed resolution, matrix size is proportional to
signal bandwidth. Moreover, as the ratio
$F_{\textrm{total}}/F_{\textrm{Landau}}$ increases, the possible
spectral support obtained at the first step of the reconstruction
increases beyond the increase of the signal bandwidth.

We note that we could reduce the run time without significantly
affecting the empirical success percentage by solving
(\ref{mat_eq2}) using a low resolution and reconstructing the signal
using a higher resolution.

The algorithm, modified as explained in the Appendix, was also
tested against real-valued signals. The assumed maximum frequency
$F_{\max}$ was set to $20$ GHz.  The number of sampling channels was
set to 3 with the sampling frequencies chosen to be $F^1=3.8$ GHz,
$F^2=4.0$ GHz, and  $F^3=4.2$ GHz resulting in a total sampling
rate, $F_{\textrm{total}}=12$ GHz. The sampling frequencies are the
same as are used in our experiments based on asynchronous MRS
\cite{Asynchronous}.  The number of bands was set to $8$ ($4$
positive and $4$ negative frequencies, assuming no carrier frequency
so low as to have the $0$ frequency in its spectrum). Each band was
chosen to be of equal width $F_{\textrm{Landau}}/8$.

Once a band $(a,b]$ was chosen, the spectrum of $X(f)$ for
$f\in(a,b]$ was determined by the following formula:
\[X(f)=A\sin\left[\frac{\pi (f-a)}{b-a}\right]e^{j\theta}.\]
The phase $\theta$ was chosen randomly from a uniform distribution
on $[0,2\pi]$ and the amplitude $A$ was chosen randomly from a
uniform distribution on $[1,1.2]$.

\begin{figure}[!t]
    \centering
        \includegraphics[width=3.5in]{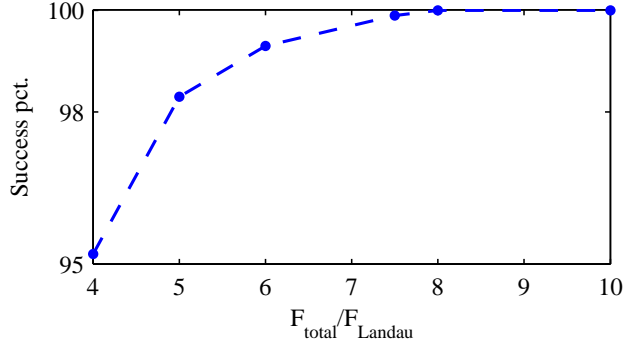}
    \caption{Empirical success percentages for equal 4-bands real signals for different spectral supports ($F_{\textrm{Landau}}$)
     with $F_{\textrm{Nyquist}}$=40 GHz,
    total sampling rate $F_{\textrm{total}}=12$ GHz }\label{fig:suc_rate1}
\end{figure}

Fig.~\ref{fig:suc_rate1} shows the empirical success percentages of
the algorithm tested against real valued signals. As is evident from
the figure, the empirical success percentage is high when
$F_{\textrm{total}}/F_{\textrm{Landau}}\ge 8$. We note that the
required sampling rate is significantly higher in this example than
in the complex signals simulation. The reason is that in our real
case example there are twice as many bands as in the complex case
simulations. Hence, after the sampling, an overlap may also occur
between the negative and the positive bands of the real signal. We
note that when sampling a real signal at a sampling rate $F^i$, it
is sufficient to know the spectrum in a frequency region
$[0,F^i/2]$. However, for real signals, there is uncertainty as to
whether a signal in baseband is obtained from a signal in the
positive band or in the negative band.

\begin{figure}[!t]
  \centering
    \includegraphics[width=3.5in]{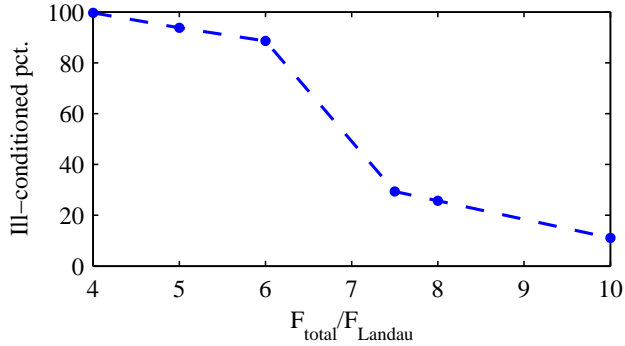}
  \caption{Ill-posed cases mean rate for 4-bands real signals for different spectral supports ($F_{\textrm{Landau}}$)
  with $F_{\textrm{Nyquist}}$=40 GHz,
    total sampling rate $F_{\textrm{total}}=12$ GHz}\label{ill_cond_real}
\end{figure}

\begin{figure}[!t]
    \centering
        \includegraphics[width=3.5in]{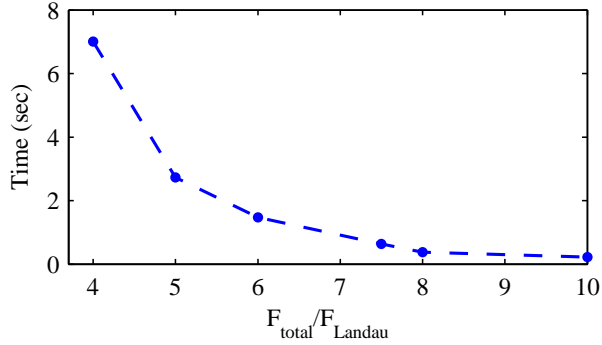}
    \caption{Mean recovery times for equal 4-bands real signals for different spectral supports (total bandwidth)
    with $F_{\textrm{Nyquist}}$=40 GHz,
    total sampling rate $F_{\textrm{total}}=12$ GHz}\label{fig:mean_time}
\end{figure}

The number of ill-posed cases and the mean recovery run times for
the real-valued signals are shown in Figs.~\ref{ill_cond_real} and
\ref{fig:mean_time} respectively. It can bee seen that the mean rate
of ill-conditioned cases is much lower for real-valued signal
simulations than for complex ones. This could be due to the
correlation between positive and negative frequency components of
real signals.

%Although the Landau rate is as much as twice as high our in real-valued signal experiments
%than in our complex ones, there is a dependency between the positive
%and negative frequencies' components. In addition the total sampling
%rate for real-valued signals is increased four times (12 GHz) with
%respect to complex-valued signals (3 GHz). For those two reason
%stated above the mean ill-posed cases rate keeps low in our
%simulations.
\label{Simulations_sect}

\subsection{Solution Stability}

\begin{figure}[!t]
    \centering
        \includegraphics[width=3.5in]{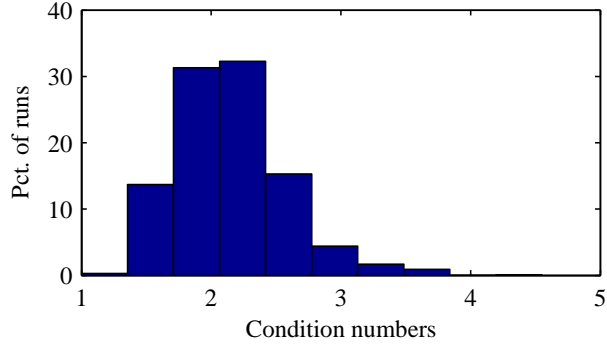}
   \caption {Condition numbers histogram of the mixing matrix in (\ref{mat_eq3_real}) for 4 bands, 200 MHz width real signals,
   $F_{\textrm{total}}=12$ GHz }
    \label{fig:cond_nums}
\end{figure}

The stability of the linear equations used in the recovery scheme
was tested via the condition number for the real-valued signals
equations ((\ref{mat_eq3_real}) in the Appendix). This case is
important in our experiments since we sample real
signals~\cite{Asynchronous}. %containing columns that correspond to the original signal frequencies.
The reconstruction scheme for real-valued signals requires solving
two systems of linear equations; one for the real part and the other
for the imaginary part. Each of the two systems of equations is
described by a different matrix. In each test case we presented the
maximum value of the condition numbers of the two matrices. For
4-band 200-MHz-width randomly generated signals
($F_{\textrm{Landau}}$=1.6 GHz) the condition number among 1000 runs
was at most 5.3. The condition numbers histogram is shown in
Fig.~\ref{fig:cond_nums}.

\begin{figure}[!t]
    \centering
        \includegraphics[width=3.5in]{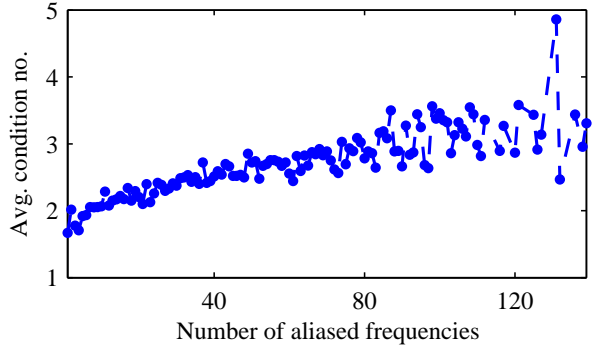}
   \caption {Condition numbers mean value vs. number of aliased frequencies for 4 bands, 200 MHz width real signals,
   $F_{\textrm{total}}=12$ GHz }
    \label{fig:cond_nums_vs_aliased}
\end{figure}

When sampling a sparse signal, most frequencies of the signal are
unaliased in at least one of the sampling channels. It can be easily
shown that when a frequency component of the signal is not aliased
in any of the sampling channels, the reconstructed signal is
obtained simply by averaging the corresponding sampled spectrum at
the different sampling channels. The reconstruction of a frequency
component that is unaliased in at least one of the sampling channels
can also be easily performed by copying the corresponding unaliased
spectrum to the reconstructed signal. Therefore, for sparse signals,
the reconstruction in the SMRS scheme is robust and the condition
number of the matrices is small. Indeed, we have verified that the
condition number of the matrix increases as the number of
frequencies in the original signal that are aliased increases. Fig.~
\ref{fig:cond_nums_vs_aliased} shows the mean values of condition
numbers versus number of aliased frequencies as calculated for the 4
real signals each with a 200 MHz width that are sampled at
$F_{\textrm{total}}=12$ GHz. The other parameters of the simulation
are the same as those in the simulation that resulted in
Fig.~\ref{fig:suc_rate1}.

\subsection{Noisy signals}

The algorithm's performance was also tested for its ability to
reconstruct real-valued signals contaminated by Gaussian white
noise. The presence of noise demands some modification of the
algorithm. One modification is in detecting the possible bands of
the support of the originating signal. Because the spectral support
of white noise is not restricted to the spectral support of the
uncontaminated signal, the indicator functions in (\ref{indicator})
cannot be used. Instead, we adapt (\ref{indicator}) to noisy cases
similarly as in ~\cite{Asynchronous}. In~\cite{Asynchronous}, for the
indicator function $\chi^i[l]$ to be equal to 1 at any frequency, it
was required that the average energy of the signal in the
neighborhood of that frequency be higher than a certain threshold.
In SMRS we further expand each band in $\chi[l]$ to include
additional frequencies that might otherwise be omitted when defining
the indicator functions $\chi^i[l]$. Once the bands are identified
the matrix equations are constructed exactly as in the noiseless
case.

%Because the spectral support of white noise affects the entire
%spectrum, a signal contaminated by white noise can no longer be
%considered a multi-band in the strict sense.  Thus one cannot expect
%to reconstruct it perfectly at any rate lower than the Nyquist rate.
%%Consequently, the system of equations~\ref{???} will be overdetermined.
%This is reflected in the fact that equation \ref{mat_eq3_real} no longer has a
%solution. Nevertheless, if the matrix
%$\underline{\underline{\widehat{A}}}^{r,im}_{\textrm{red}}$ is full column
%rank, the pseudo-inverse can still be used.
The solution of the linear equations given in (\ref{mat_eq3_real})
is modified in the noisy case. Because the added white noise affects
the entire spectrum, a signal contaminated by white noise can no
longer be considered multiband in the strict sense. Thus one cannot
expect to reconstruct it perfectly from samples taken at a total
rate lower than the Nyquist rate. Whereas in the ideal noiseless
case the error norm vanishes, with a signal containing noise, one
must relent on a perfect reconstruction and settle for a minimum
error. In the noisy case the solution to (\ref{mat_eq3_real}) should
solve the least square problem
$\min_{\textbf{x}^{\textrm{r,im}}_{\textrm{red}}}\|\widehat{\textbf{x}}^{\textrm{r,im}}_{\textrm{red}}-
\widehat{\textbf{A}}^{\textrm{r,im}}_{\textrm{red}}~
\textbf{x}^{\textrm{r,im}}_{\textrm{red}}\|$. When the the matrix
$\widehat{\textbf{A}}^{\textrm{r,im}}_{\textrm{red}}$ is not full
column rank, we use the modified OMP algorithm which is adjusted to
account for the errors due to noise. As noted above, in the
noiseless case, one can expect a perfect reconstruction and thus the
threshold error $\epsilon$ can be set to $0$ or a very small number.
However, with noisy signals, some care must be taken in choosing
$\epsilon$. On the one hand, if the $\epsilon$ is chosen too large,
the algorithm may stop before a solution is reached. On the other
hand, if $\epsilon$ is chosen too small, the reconstructed signal
may include bands that are not in the originating signal. The
problem of too high threshold is solved by changing the stop
criterion. Instead of stopping the algorithm when a threshold is
attained, we check at each iteration whether the block that reduces
the residual error the most causes the resulting matrix to be rank
deficient. When this occurs, the iteration are stopped and the block
is not added to the matrix.

An additional change is made to the algorithm when treating the
blocks. In the noiseless case each block corresponds to a single
band in $\chi[l]$. When sampling noisy signals, we divide each block
into several sub-blocks. The reason for this division is that, with
noisy signals, the identification of the bands is not accurate.
Identified bands may be wider than the originating bands. This is
particularly true if the threshold is chosen small.  This widening
may cause the inclusion of false frequencies whose corresponding
columns in $\widehat{\textbf{A}}^{\textrm{r,im}}_{\textrm{red}}$ are
linearly dependent on the columns corresponding to the support of
the originating signal. By using smaller sub-blocks such columns may
be isolated from the rest of the columns in their corresponding
band.

The recovery scheme was tested against real-valued signals with $8$
bands ($4$ positive frequencies bands and $4$ negative frequencies
bands). The signals without noise were generated and sampled exactly
as in the noiseless simulations of real signals. Noise was added
randomly at each frequency of the pre-sampled signal according to a
normal distribution with standard deviation $\sigma=0.04$; the SNR
was defined by $10\log_{10}(1/(\sigma\sqrt{F_{\max}/F^2}))=10.5$ dB.
This definition takes into account the accumulation of noise in
baseband due to sampling. The sampling rates were the same as those
in the noiseless simulations. The indicator functions $\chi^i[l]$
were constructed using the same parameters as those used in
\cite{Asynchronous}. Each band in $\chi[l]$ was widened by 20
percent on each side. The sub-blocks used in the modified OMP had
spectral width of 100 MHz.
\begin{figure}[!t]
    \centering
        \includegraphics[width=3.5in]{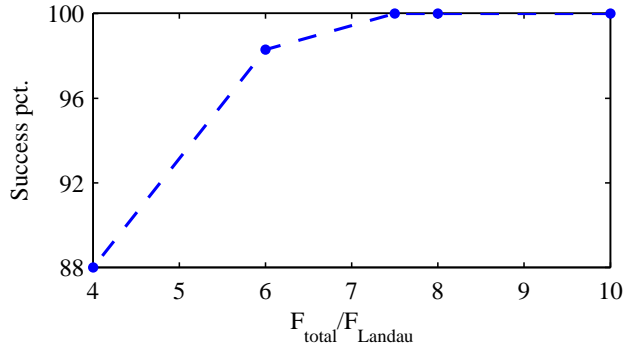}
   \caption {Empirical success percentages for 4 bands of real signals noise that were contaminated with a noise with a standard deviation of $\sigma=0.04$ for different spectral supports of the signal
   ($F_{\textrm{Landau}}$) with $F_{\textrm{Nyquist}}$=40 GHz,
    and a total sampling rate $F_{\textrm{total}}=12$ GHz }\label{fig:suc_rate_noise}
\end{figure}
The success was measured by the algorithm's ability to achieve a low
error $l_1$ norm below $2\sigma\sqrt{F_{\max}/F^2}=4.47\sigma$ for each recovered %EXPLAIN
band. The mean error for each recovered band $X_\textrm{rec}(f)$ and
the true band $X(f)$ were evaluated as follows:
\[\frac{1}{|B|}\int_B |X_\textrm{rec}(f)-X(f)|df<2\sigma\sqrt{F_{\max}/F^2}\] where $B$ is the
band support.

Statistics on recovering $8$ bands 200 MHz width each are based on
10000 tests. The simulation showed that, although the algorithm's
performance inevitably decreased, it still achieved a high empirical
recovery rate (37 failures out of 10000 tests). Additional
simulations were performed by changing the Landau rates as was done
in the simulations performed for the noiseless case. In
Fig.~\ref{fig:suc_rate_noise} the empirical success percentage is
presented for 1000 simulations of noisy signals. The results of the
simulations are similar to those in the noiseless case. When the
total sampling rate is 8 times higher than the Landau rate, high
success percentage was achieved. The recovery error level depended
on the threshold choice. Lower threshold allows more accurate
reconstruction but increases the recovery time. Moreover, additional
parameters adjustments are also necessary (widening percentage,
sub-blocks size). Different error criteria are also possible. For
example, choosing $l_2$ norm instead of $l_1$ norm and setting the
error threshold to be $3.3\sigma$ as in \cite{Asynchronous} resulted
in $99.5$ percent empirical success rate in recovering 1.6 GHz
Landau rate signals and $99.8$ percent for 1.5 GHz Landau rate
signals. This is in contrast to simulations results in
Fig.~\ref{fig:suc_rate_noise} where empirical perfect reconstruction
was obtained for those signals under different criteria.

\section{Conclusions}

In this paper we describe a synchronized multirate sampling scheme
for accurately reconstructing sparse multiband signals using a small
number of sampling channels (3 in our simulations) whose total
sampling rate is significantly lower than the Nyquist rate.

Although the same data that is obtained from our scheme can be
obtained by a multicoset scheme, such a multicoset scheme requires
many more channels and a time accuracy that cannot be attained
practically.   Moreover, our scheme processes the data differently,
in a way that results in significantly wider basebands and thus
greatly reduces the effects of aliasing.  By synchronizing the
sampling channels our scheme is able to reconstruct signals
correctly even in most cases in which the signal is aliased in all
the sampling channels.

The main advantage of multicoset sampling schemes is theoretical. Because in a
multicoset sampling scheme each channel samples at the same frequency, one has
a mathematical structure that enables a perfect reconstructions of ideal
multiband signals from samples taken at a total sampling rate that is close to
a theoretical lower bound. This bound is attained only under special
assumptions regarding the number and width of the signal bands. Moreover, the
bound requires the number of sampling channels to be twice the number of signal
bands \cite{Mishali}. Hence, in many cases the number of sampling channels
becomes too high for a practical implementation.

The main advantage of the SMRS is in the use of a small number of
sampling channels that operate at relatively high rates that can
reconstruct accurately sparse signal that consist of several bands.
Sampling at higher rates has a fundamental advantage in that it
increases the SNR after sampling. Another advantage is that
implementation of the SMRS scheme does not require a priori
knowledge of the maximum width of the signal bands. A third
advantage of our scheme is that, in many cases the signal can be
reconstructed by simple matrix inversion rather than through a
search algorithm as in the multicoset recovery scheme of
\cite{Mishali}.

In the SMRS scheme, when sampling sparse signals, most of the
sampled spectrum is unaliased in at least one of the sampling
channels. Hence, most of the spectrum can be reconstructed directly
from the unaliased parts of the spectrum. On the other hand, in
multicoset sampling schemes an alias in one channel is equivalent to
an alias in all channels. Furthermore, a multicoset scheme
downconverts signals to much lower frequencies, thus increasing the
negative effects of aliasing. Our numerical simulations indicate
that the reduced aliasing in our SMRS scheme results in
significantly better performance over a multicoset sampling scheme
of \cite{Mishali} whose number of sampling channels is small. Also,
due to reduced aliasing, it is expected that the reconstruction in
the SMRS scheme will be robust when noise is added in the sampling
process.

Although we do not have a rigorous criteria for perfect
reconstruction, by examining the multicoset pattern that yields the
same data it might be possible to obtain necessary conditions for a
perfect reconstruction.

Our scheme also solves ill-conditioned linear equations by using a
modified OMP algorithm. Whereas we obtained satisfactory results
with it, we do not have criteria for determining when the modified
OMP algorithm converges to the correct solution. However, this
shortcoming may be not as significant because of the possibility of
using other algorithms to solve the equations.

\section{Appendix}
In this appendix we present the modifications to (\ref{mat_eq2}) for
the real signals recovery. Since the signal is real-valued, its
spectrum fulfills
\begin{equation}\label{conjugate}
X(f)=\overline{X}(-f)
\end{equation}
where $\overline{a+bj}=a-bj$ is the complex conjugate and $a$ and $b$ are real numbers.

It follows from (\ref{conjugate}) and (\ref{sampled spectrum}), that
for each channel index $i$, all the information about $X^i(f)$ is
contained in the interval $[0,F^i/2]$. Consequently, it is
convenient to choose the sampling frequencies $F^i$ such that
$F^i/2=\Delta f M^i/2$ where $M^i/2$ is an integer. Because the
conjugation operation $\overline{a+jb}: a+jb \mapsto a-jb$ is not
complex linear, (\ref{mat_eq}) needs to be replaced with
two systems of equations; one for the real part and one for the
imaginary part.

We use the following notations to represent the spectrum of the real signals in the discretized frequencies:
\begin{align}\label{discrete_real}
&X^i[k]=X^i(k\Delta f)~~~~k=-\lfloor M^i/2\rfloor,  \ldots, \lfloor M^i/2\rfloor,\\
\notag &X[k]=X(k\Delta f) ~~~~k=-\lfloor M/2\rfloor,  \ldots,
\lfloor M/2 \rfloor.
\end{align}

The sequence $X^i[k]$ contains the samples of $X^i(f)$ in the
baseband $[-F^i/2,F^i/2]$. The sequence $X[k]$ contains the samples
of $X(f)$ given in $[-M\Delta f/2,M\Delta f/2 ]$, where $M$ is
chosen to fulfill $M=\lceil F_{\textrm{Nyquist}}/\Delta f\rceil$.  %CHECK>>>>>>>>>>>>>>>>>>>>>>>>
Equation (\ref{baseband sum}) now takes the following form:
\begin{equation}\label{sum_delt_real}
    X^i[k]=F^i\sum_{l=-\lfloor M/2\rfloor}^{\lfloor M/2\rfloor}X[l]\sum^{\infty}_{n=-\infty}\delta[l-(k+nM^i)].
\end{equation}
Equation (\ref{sum_delt_real}) can be written in a matrix form as
\begin{equation} \label{mat_eq_real}
\textbf{x}^i=\textbf{A}^i \textbf{x}
\end{equation}
where $\textbf{x}^i$ and $\textbf{x}$ are given by
\begin{align}\label{elements1_real}
(\textbf{x}^i)_{k+\lfloor M^i/2\rfloor+1}&=X^i[k],~~-\lfloor M^i/2\rfloor\leq k\leq \lfloor M^i/2\rfloor,\\
\notag (\textbf{x})_{k+\lfloor M/2\rfloor+1}&=X[k],~~-\lfloor
M/2\rfloor\leq k\leq \lfloor M/2\rfloor,
\end{align}
and $\textbf{A}^i$ is a matrix whose elements are given by
\begin{equation}\label{basis_func_real}
\textbf{A}_{k+\lfloor M^i/2\rfloor+1,l+\lfloor M/2\rfloor+1}^i=F^i
\sum^{\infty}_{n=-\infty}\delta[l-(k+nM^i)].
\end{equation}
Note that, since the signal is real valued, all of its spectral information is contained
in the positive frequencies.

Each element in $\textbf{A}^i$ is equal to either $F^i$ or 0.
Equation (\ref{mat_eq_real}) for the different sampling channels can
be concatenated as in complex signals case to yield
\begin{equation} \label{mat_eq2_real}
\widehat{\textbf{x}}=\widehat{\textbf{A}}\textbf{x}.
\end{equation}
%The real and
%imaginary parts of the signal's spectrum are linearly dependent because of Eq.~\ref{conjugate}:
%\begin{align}
%&\mathrm{Re}(\underline{X}_k)=\mathrm{Re}(\underline{X}_{2M+2-k}) ~~~~k=1,\dots,2M+1 \\
%\notag
%&\mathrm{Im}(\underline{X}_k)=-\mathrm{Im}(\underline{X}_{2M+2-k}).
%\end{align}
%This dependency can be inserted into Eq.~\ref{mat_eq2_real} by
%decomposing it to its real and imaginary parts:
The spectrum can be decomposed into its real and imaginary parts. As
a result (\ref{mat_eq2_real}) becomes
\begin{align} \label{mat_eq3_real}
&\widehat{\textbf{x}}^r=\widehat{\textbf{A}}^r~ \textbf{x}^r,\\
\notag & \widehat{\textbf{x}}^{im}=\widehat{\textbf{A}}^{im}
\textbf{x}^{im}
\end{align}
where $\widehat{\textbf{x}}^r=\mathrm{Re}(\widehat{\textbf{x}})$ and
$\widehat{\textbf{x}}^{im}=\mathrm{Im}(\widehat{\textbf{x}})$. In
addition only components that correspond to positive frequencies are
retained in the vectors $\widehat{\textbf{x}}^r$ and
$\widehat{\textbf{x}}^{im}$.
% the
%elements of $\widehat{\textbf{x}}^r$ and $\widehat{\textbf{x}}^{im}$
%are given by
%\begin{align}
%&(\widehat{\textbf{x}})_l^r=\mathrm{Re}((\textbf{x})_{M+l}),~~l=1,\ldots,M+1 \\
%\notag
%&(\widehat{\textbf{x}})_l^{im}=\mathrm{Im}((\textbf{x})_{M+l}),~~l=1,\ldots,M+1,
%\end{align}
The elements of the matrices $\widehat{\textbf{A}}^r$ and
$\widehat{\textbf{A}}^{im}$  are given by
\begin{align}
&\widehat{\textbf{A}}_{k,\lfloor
M/2\rfloor-l+1}^r=\widehat{\textbf{A}}_{k,l+1}+
\widehat{\textbf{A}}_{k,M-l},~~l=0,\ldots,\lfloor M/2\rfloor,\\
%\notag &\widehat{\textbf{A}}_{k,1}^r=\widehat{\textbf{A}}_{k,M+1} \\
\notag &\widehat{\textbf{A}}_{k,\lfloor
M/2\rfloor-l+1}^{im}=\widehat{\textbf{A}}_{k,M-l}-\widehat{\textbf{A}}_{k,l+1},~~l=0,\ldots,\lfloor
M/2\rfloor.
\end{align}
%Equation (\ref{mat_eq3_real}) can be also rewritten as a single
%matrix equation as follows
%\begin{equation}\label{mat_eqfull}
%\left(
%\begin{array}{c}
%\widehat{\textbf{x}}^r\\
%\widehat{\textbf{x}}^{im}
%\end{array}
%\right) =\left(\begin{array}{cc}
%\widehat{\textbf{A}}^r&0\\
%0&\widehat{\textbf{A}}^{im}
%\end{array}
%\right) \left(
%\begin{array}{c}
%\textbf{x}^r\\
%\textbf{x}^{im}
%\end{array}
%\right) .
%\end{equation}
%
The reconstruction is performed with (\ref{mat_eq3_real}) exactly as
in the complex case.


\begin{thebibliography}{1}
\bibitem {Landau}
H. Landau, "Necessary density conditions for sampling and
interpolation of certain entire functions," \textit{Acta Math.},
vol. 117, pp. 37-52, July 1967.

\bibitem{Kohlenberg}
A. Kohlenberg, "Exact Interpolation of Band-limited Functions,"
\textit{Appl. Phys.}, vol. 24,pp. 1432-1436, 1953.

\bibitem{Venkantaramani}
R. Venkantaramani and Y. Bresler, "Optimal sub-Nyquist nonuniform
sampling and reconstruction for multiband signals," \textit{IEEE
Trans. Signal Process.}, vol. 49, pp. 2301-2313, Oct. 2001.

\bibitem{Lu}
Y. M. Lu and M. N. Do, "A Theory for Sampling Signals from a union
of Sub-spaces," \textit{IEEE Trans. Signal Process.}, to be
published.

\bibitem {Mishali}
M. Mishali and Y. Eldar, "Blind multiband signal recostruction:
compressed sensing for analog signals," \textit{arXiv:0709.1563},
Sept. 2007.

\bibitem{Lin}
Y. P. Lin and P. P. Vaidyanathan,  "Periodically uniform sampling of
bandpass signals," \textit{IEEE Trans. Circuits Sys.}, vol. 45, pp.
340-351, Mar. 1998.
\bibitem{Herley}
C. Herley and W. Wong, "Minimum rate sampling and reconstruction of
signals with arbitrary frequency support," \textit{IEEE Trans.
Inform. Theory}, vol. 45, pp. 1555-1564, Jul. 1999.

\bibitem{Elad}
A.M.~Bruckstein, D.L.~Donoho and M.~Elad, "From Sparse Solutions of
Systems of Equations to Sparse Modeling of Signals and Images,"
\textit{SIAM}, to be published.

\bibitem{Asynchronous}
A.~Rosenthal, A.~Linden and M.~Horowitz, "Multirate asynchronous
sampling of multiband signals," submitted for publication. % to Opt.~Soc.~Am.~B.

\bibitem{Avi}
A.~Zeitouny, A.~Feldser, and M.~Horowitz, ``Optical sampling of
narrowband microwave signals using pulses generated by
electroabsorption modulators,'' \textit{Opt. Comm.}, vol. 256, pp.
248-255, Dec. 2005.

\bibitem{Stewart}
I. Stewart and D. Tall, \textit{The Foundations of Mathematics},
Oxford, England: Oxford University Press, 1977.

\bibitem{Penrose}
R.~Penrose,"A generalized inverse for matrices," in \textit{Proc.
Cambridge Philosophical Society}, Cambridge, vol. 51, 1955, pp.
406-413.

\bibitem{Bresler2}
R.~Venkataramani and Y.~Bresler, "Sampling theorems for uniform and
periodic nonuniform MIMO sampling of multiband signals,"
\textit{IEEE Trans. Signal Process.}, vol. 51, pp. 3152-3163, Dec.
2003.

\bibitem {Candes}
E. J. Cand`es, J. Romberg and T. Tao, "Robust uncertainty
principles: exact signal reconstruction from highly incomplete
frequency information," \textit{IEEE Trans. Inform. Theory}, vol.
52, pp. 489-509.


\end{thebibliography}
\end{document}